\documentclass[12pt]{article}
\usepackage{graphicx}
\begin{document}
\begin{center}
{\large\bf Unitary and Asymptotic Behavior of Amplitudes in
Non-Anticommutative Quantum Field Theory}
\vskip 0.3 true in {\large J. W. Moffat}
\vskip 0.3 true in
{\it Department of Physics, University of Toronto, Toronto,
Ontario M5S 1A7, Canada}
\end{center}
\begin{abstract}%
The unitarity condition for scattering amplitudes in a non-anticommutative quantum
field theory is investigated. The Cutkosky rules are shown to hold for Feynman
diagrams in Euclidean space and unitarity of amplitudes can be satisfied. An analytic
continuation of the diagrams to physical Minkowski spacetime can be performed without
invoking unphysical singularities in amplitudes. The high energy behavior
of amplitudes is found to be regular at infinity provided that only space-space
non-anticommutativity is allowed.
\end{abstract}



\section{Introduction}

In previous work~\cite{Moffat}, we developed a superspace
fomalism based on coordinates
\begin{equation}
\label{supercoordinates} \rho^\mu=x^\mu+\beta^\mu, \end{equation}
where $x^\mu$ denote the familiar spacetime coordinates saisfying
\begin{equation}
[x^\mu,x^\nu]=0,
\end{equation} and $\beta^\mu$
are coordinates in an associative Grassman algebra which satisfy
\begin{equation}
\{\beta^\mu,\beta^\nu\}\equiv\beta^\mu\beta^\nu+\beta^\nu\beta^\mu=0.
\end{equation} Here, $\mu=0...3$ although the formalism can
easily be extended to higher-dimensional spaces.

The product of two operators ${\hat\phi}_1$ and ${\hat\phi}_2$
in our superspace is given by the $\circ$-product
\begin{equation}
\label{phiproduct}
({\hat\phi}_1\circ{\hat\phi}_2)(\rho)=\biggl[\exp\biggl(\frac{1}{2}\omega^{\mu\nu}
\frac{\partial}{\partial\rho^\mu}\frac{\partial}{\partial\eta^\nu}\biggr)
\phi_1(\rho)\phi_2(\eta)\biggr]_{\rho=\eta} $$ $$
=\phi_1(\rho)\phi_2(\rho)+\frac{1}{2}\omega^{\mu\nu}\frac{\partial}{\partial\rho^\mu}
\phi_1(\rho)\frac{\partial}{\partial\rho^\nu}\phi_2(\rho)+O(\omega^2),
\end{equation}
where $\omega^{\mu\nu}$ is a nonsymmetric tensor
\begin{equation} \omega^{\mu\nu}=-\tau^{\mu\nu}+i\theta^{\mu\nu},
\end{equation}
with $\tau^{\mu\nu}=\tau^{\nu\mu}$ and
$\theta^{\mu\nu}=-\theta^{\nu\mu}$. Moreover, $\omega^{\mu\nu}$ is
Hermitian symmetric $\omega^{\mu\nu}=\omega^{\dagger\mu\nu}$, where
$\dagger$ denotes Hermitian conjugation.

Let us define the notation
\begin{equation}
[\phi_1(\rho),\phi_2(\rho)]_\circ\equiv
\phi_1(\rho)\circ\phi_2(\rho)-\phi_2(\rho)\circ\phi_1(\rho),
\end{equation}
and
\begin{equation}
\{\phi_1(\rho),\phi_2(\rho)\}_\circ\equiv
\phi_1(\rho)\circ\phi_2(\rho)+\phi_2(\rho)\circ\phi_1(\rho).
\end{equation}
From (\ref{supercoordinates}) and (\ref{phiproduct}), we obtain for the
superspace operator ${\hat\rho}$:
\begin{equation}
\label{circnoncom}
[{\hat\rho}^\mu,{\hat\rho}^\nu]_\circ=2\beta^\mu\beta^\nu+i\theta^{\mu\nu}+O(\theta^2),
\end{equation}
\begin{equation} \label{circnonanticom}
\{{\hat\rho}^\mu,{\hat\rho}^\nu\}_\circ=2x^\mu
x^\nu+2(x^\mu\beta^\nu+x^\nu\beta^\mu) -\tau^{\mu\nu}+O(\tau^2).
\end{equation}

If we now choose the limit in which $\beta^\mu\rightarrow 0$ and
$\tau^{\mu\nu}\rightarrow 0$, then we get from (\ref{circnoncom}) and
(\ref{circnonanticom}):
\begin{equation}
\label{standardnoncom}
[{\hat\rho}^\mu,{\hat\rho}^\nu]_\circ\rightarrow [{\hat x}^\mu,{\hat
x}^\nu]=i\theta^{\mu\nu},
\end{equation}
\begin{equation}
\label{standardanti}
\{{\hat\rho}^\mu,{\hat\rho}^\nu\}_\circ\rightarrow \{{\hat x}^\mu,{\hat
x}^\nu\}=2x^\mu x^\nu.
\end{equation}
We see that (\ref{standardnoncom}) and (\ref{standardanti})
give us back the usual noncommutative expressions for the coordinate
operators ${\hat x}^\mu$.

In the limit $x^\mu\rightarrow 0$ and $\theta^{\mu\nu}\rightarrow 0$, we get from
(\ref{circnoncom}) and (\ref{circnonanticom}):
\begin{equation}
[{\hat\rho}^\mu,{\hat\rho}^\nu]_\circ\rightarrow
[{\hat\beta}^\mu,{\hat\beta}^\nu]=2\beta^\mu\beta^\nu,
\end{equation}
\begin{equation}
\label{rhoequation}
\{{\hat\rho}^\mu,{\hat\rho}^\nu\}_\circ\rightarrow
\{{\hat\beta}^\mu,{\hat\beta}^\nu\}=-\tau^{\mu\nu}.
\end{equation}

In the following, we shall the consider the simpler geometry
determined by $\theta^{\mu\nu}=0$. We define a
$\diamondsuit$-product
\begin{equation}
\label{diamondproduct}
({\hat\phi}_1\diamondsuit{\hat\phi}_2)(\rho)=\biggl[\exp\biggl(-\frac{1}{2}\tau^{\mu\nu}
\frac{\partial}{\partial\rho^\mu}\frac{\partial}{\partial\eta^\nu}\biggr)
\phi_1(\rho)\phi_2(\eta)\biggr]_{\rho=\eta} $$ $$
=\phi_1(\rho)\phi_2(\rho)-\frac{1}{2}\tau^{\mu\nu}\frac{\partial}{\partial\rho^\mu}
\phi_1(\rho)\frac{\partial}{\partial\rho^\nu}\phi_2(\rho)+O(\tau^2).
\end{equation} We now have \begin{equation}
\{{\hat{\rho}^\mu,{\hat\rho}^\nu}\}_\diamondsuit=2x^\mu
x^\nu+2(x^\mu\beta^\nu+x^\nu\beta^\mu)-\tau^{\mu\nu}.
\end{equation}

A calculation of the one loop diagram in scalar field theory
is finite, and the higher order loops will be
finite to all orders of perturbation theory, due to the
convergence of the modified Feynman propagator ${\bar\Delta}_F$ in
momentum space~\cite{Moffat}.

We assume that there exists a classical limit as
$\ell\rightarrow 0$, $\beta^\mu\rightarrow 0$ and $\tau^{\mu\nu}\rightarrow 0$, where
$\ell$ is a natural unit of length, so that we obtain the
classical c-number spacetime continuum with $\{{\hat x}^\mu,{\hat x}^\nu\}=2x^\mu
x^\nu$.

Unitarity of the S-matrix and crossing symmetry are two of the most important
features of quantum field theory. In the following, we shall explore the
consequences of unitarity for amplitudes in non-anticommutative scalar field theory,
the positions of singularities in the amplitudes and the asymptotic properties of
amplitudes. We consider the case when $\tau^{mn}\not= 0$ and
$\tau^{00}=\tau^{0n}=0\quad(m,n=1,2,3)$, because only then can we
define a conjugate momentum operator and avoid difficulties with
acausal dynamics. Moreover, we shall find for this case that the
amplitudes have regular behavior at infinity.

\section{\bf Non-anticommutative Field Theory}

We have for
two operators ${\hat f}$ and ${\hat g}$ in our superspace manifold:
\begin{equation}
({\hat f}\diamondsuit {\hat g})(\rho)=\frac{1}{(2\pi)^8}\int d^4k
d^4q{\tilde f}(k){\tilde g}(q)\exp\biggl[\frac{1}{2}(k\tau
q)\biggr]\exp[i(k+q)\rho],
\end{equation}
where
\begin{equation}
{\tilde f}(k)=\frac{1}{(2\pi)^4}\int d^4\rho\exp(-ik\rho)f(\rho),
\end{equation}
and $(k\tau q)\equiv k_\mu\tau^{\mu\nu}q_\nu$.
We shall assume that derivatives act trivially
in our superspace
\begin{equation}
[\rho_\mu,\partial_\nu]=-\eta_{\mu\nu},\quad
[\partial_\mu,\partial_\nu]=0,
\end{equation}
where $\partial_\mu=\partial/\partial\rho_\mu$.

In contrast to the noncommutative field theories, the kinetic energy
component of the action in non-anticommutative field theories is not
trivially the same as the commutative field theories, because of the
symmetry of the tensor $\tau^{\mu\nu}$. The action for the scalar field
$\phi$ is
\begin{equation}
I=\int
d^4\rho\biggl[\frac{1}{2}\partial_\mu\phi(\rho)\diamondsuit\partial^\mu\phi(\rho)
-\frac{m^2}{2}\phi(\rho)\diamondsuit\phi(\rho)-V_\diamondsuit(\phi)\biggr],
\end{equation} where $V_\diamondsuit(\phi)$ is the scalar field
potential. For our calculations we shall choose the potential:
\begin{equation}
\label{potential}
V_\diamondsuit(\phi)=\frac{\lambda}{4!}\phi^\beta_\diamondsuit,
\end{equation}
where $\phi^\beta_\diamondsuit$ denotes the
$\beta$th order product of the field $\phi$ using the
$\diamondsuit$-product. The commutative theories $\beta=3$ and
$\beta=4$ give the well-known renormalizable theories, while for
$\beta\ge5$ the commutative theory is non-renormalizable.

In the noncommutative and the non-anticommutative cases,
there is an ambiguity in applying the quantization procedure in position space. The
usual quantization conditions are defined for the scalar field $\phi$ and its
conjugate momentum $\pi$ at different spacetime points, while the $\star$ and
$\diamondsuit$-products only make sense when the products are computed at the same
spacetime point. We can avoid this problem by working only in momentum space when
performing Feynman rule calculations to obtain matrix elements. We formally define a
$\bigtriangleup$-product of field operators ${\hat f}(\rho)$ and
${\hat g}(\eta)$ at different superspace points $\rho$ and
$\eta$:
\begin{equation}
{\hat f}(\rho)\bigtriangleup {\hat g}(\eta)
=\exp\biggl(-\frac{1}{2}\tau^{\mu\nu}\frac{\partial}{\partial
\rho^\mu}\frac{\partial}{\partial \eta^\nu}\biggl)f(\rho)g(\eta).
\end{equation}
This product reduces to the $\diamondsuit$-product
of ${\hat f}$ and ${\hat g}$ in the limit $\rho\rightarrow \eta$.

Let us now consider the Feynman rules for the non-anticommutative
scalar field theory in our superspace with coordinates
$\rho^\mu$. The modified Feynman propagator ${\bar\Delta}_F$ is
defined by the vacuum expectation value of the time-ordered
$\bigtriangleup$-product~\cite{Moffat}.
\begin{equation}
i{\bar\Delta}_F(\rho-\eta)\equiv\langle 0\vert
T(\phi(\rho)\bigtriangleup\phi(\eta))\vert 0 \rangle $$ $$
=\frac{i}{(2\pi)^4}\int\frac{d^4k\exp[-ik(\rho-\eta)]\exp[\frac{1}{2}(k\tau
k)]}{k^2-m^2+i\epsilon}.
\end{equation}
In momentum space this
gives
\begin{equation}
i{\bar\Delta}_F(k)=\frac{i\exp[\frac{1}{2}(k\tau k)]}
{k^2-m^2+i\epsilon},
\end{equation}
which reduces to the standard
commutative field theory form for the Feynman propagator
\begin{equation}
i\Delta_F(k)=\frac{i}{k^2-m^2+i\epsilon}
\end{equation} in the limit $\vert\tau^{\mu\nu}\vert\rightarrow
0$.
 
From the interaction part of the action for the theory with $\beta=4$:
\begin{equation}
I_{\rm int}=\frac{\lambda}{4!}\int
d^4\rho\phi\diamondsuit\phi\diamondsuit\phi\diamondsuit\phi(\rho),
\end{equation}
we can deduce that the vertex
factor for the scalar
non-anticommutative field theory has the form
\begin{equation}
V(k_1,k_2,k_3,k_4)=\exp\biggl\{\frac{1}{2}[(k_1\tau k_2)
+(k_2\tau k_3)+(k_3\tau k_4)]\biggr\}.
\end{equation}
The vertex
function factor for a general diagram is given by
\begin{equation}
V(k_1,...,k_n)=\sum_{i<j}\exp\biggl[\frac{1}{2}(k_i\cdot k_j)\biggr],
\end{equation}
where $k_i\cdot k_j=k_{i\mu}\tau^{\mu\nu}k_{j\nu}$.

Our Feynman rules are: for every internal line we insert a
modified Feynman propagator ${\bar\Delta}_F(k)$ and integrate
over $k$ with the appropriate numerical factor.
We associate with every diagram a vertex
factor $V(p_1,...,p_n; k_1,...,k_n)$ where the $ps$ and $ks$
denote the external and internal momenta of the diagram,
respectively.

In the non-anticommutative scalar field theory, the one loop self-energy
diagram is {\it finite} for a generic value of the external
momentum and for fixed finite values of an energy scale parameter
$\Lambda$ in Euclidean momentum space. The same holds true for the vertex one loop
corrections. The convergence of both planar and non-planar loop
diagrams should hold to all orders of perturbation theory, because of the strong
convergence of the modified Feynman propagator ${\bar\Delta}_F$ in Euclidean
momentum space.

\section{Unitarity of Amplitudes}

A basic feature of quantum field theory is unitarity, which states that the S-matrix
must satisfy the condition $SS^\dagger=1$. In terms of the $T$ matrix defined by
$S=1+iT$, we have
\begin{equation}
\frac{1}{2i}(T_{fi}-T^\dagger_{if})=\frac{1}{2}\sum_nT^\dagger_{nf}T_{ni}.
\end{equation}
Inserting momentum conservation
\begin{equation}
\langle f\vert T\vert i\rangle=(2\pi)^4\delta^4(P_f-P_i)
{\cal T}_{fi}
\end{equation}
we obtain
\begin{equation}
\frac{1}{2i}({\cal T}_{fi}-{\cal
T}^\dagger_{if})=\frac{1}{2}\sum_n(2\pi)^4\delta^4(P_n-P_i){\cal T}^\dagger_{nf}{\cal
T}_{ni}.
\end{equation}

In general, a scattering amplitude in non-anticommutative scalar field theory will
involve the products of the modified Feynman propagator ${\bar\Delta}_f(k^2)$ for
each internal line of a diagram with Euclidean momentum $k$ and vertex functions
$V(k_1,...,k_n; p_1,...,p_n)$. Then, the amplitude describing a given diagram
with $n$ external Euclidean momenta $q_j$, which satisfies the conservation law
$q_1+q_2+...+q_n=0$, will be of the form
\begin{equation}
A=\int...\int\prod_i
d^4k_i\prod_j\frac{W_j(k_j^2,k_jq_j)}{k_j^2+m_j^2},
\end{equation}
where $k_j$ is an Euclidean
4-momentum and $m_j$ is the mass of the jth particle. The integration is carried out
over the internal line momentum $k_j$. Moreover, $W(k^2_j,k_jq_j)$ is an {\it entire}
function in the complex $k$ plane and decreases rapidly as ${\rm
Re}k\rightarrow +\infty$. For our non-anticommutative field theory, the
asymptotic behaviour $k_j^2\rightarrow\infty$ is
\begin{equation}
W_j(k_j^2,k_jq_j)\sim \exp(-ak_j^2/\Lambda^2),
\end{equation}
where $a$ is a constant. When the natural unit of length
$\ell\rightarrow 0$, yielding the classical spacetime continuum,
it is understood that this corresponds to
$\Lambda\rightarrow\infty$.

Apart from a constant
factor, the Euclidean amplitude $A$ should coincide with the real amplitude
corresponding to a region with $p_ip_j=-q_iq_j$, where the $p_i$ and $q_i$ denote the
momenta in the physical region and in Euclidean momentum space, respectively. By an
analytic continuation, we can obtain the amplitude in the physical region of the
invariant external momenta $p_i$ with the understanding that the masses involve an
additional negative imaginary part: $m_i\rightarrow m_i-i\epsilon$. When
$W(k_i,q_j)=1$ or is just a polynomial with respect to $k$, then $A$ coincides
with the real Minkowski amplitude of standard local field theory, and the Minkowski
and Euclidean formulations are completely equivalent~\cite{Schwinger}.

The analysis by Landau~\cite{Landau} of the
singularities of amplitudes in perturbation theory is based on the amplitudes in
Euclidean space. In the non-anticommutative field theories, we must determine the
singularities of the amplitude with respect to the invariant momentum variables.
We can rewrite the amplitude using the Feynman parameterization
\begin{equation}
A=(N-1)!\int_0^1...\int_0^1d\alpha_1...d\alpha_n\delta(1-\sum_{i=1}^n\alpha_i)
\int...
$$ $$
\times\int\prod_id^4k_i\prod_j\frac{W_j(k_j^2)}{\sum_j\alpha_j(k_j^2+m_j^2)},
\end{equation}
where $N$ is the number of internal lines. Following Landau~\cite{Landau}, we obtain
\begin{equation}
\sum_{i=1}^n\alpha_i(k_i^2+m_i^2)=\chi(\alpha,q_iq_j,m_j^2)+K(\alpha,k'),
\end{equation}
where $\chi$ is a
non-homogeneous quadratic form in the momenta $q_i$ that describes the free ends of
the diagram, and $K$ is a homogeneous quadratic form in the integration variable
$k'$ with coefficients that only depend on the parameters $\alpha_i$.

Since the numerator of the amplitude contains an entire function of the scalar
products $q_iq_j$ and the parameters $\alpha_i$, it cannot lead to any additional
singularities in a finite region of the invariant momentum variables. However, to
determine whether our amplitude satisfies the Cutkosky cutting rules~\cite{Cutkosky},
and thereby the unitarity relation for the S-matrix, we must calculate the
discontinuities of $A$ across the appropriate branch cuts.

We decompose the amplitude $A$ into two parts $A_I$ and $A_{II}$ connected by $r$
internal lines as shown in Fig 1.
\vskip 0.3 in
\begin{center}\includegraphics[width=3in,height=1in]{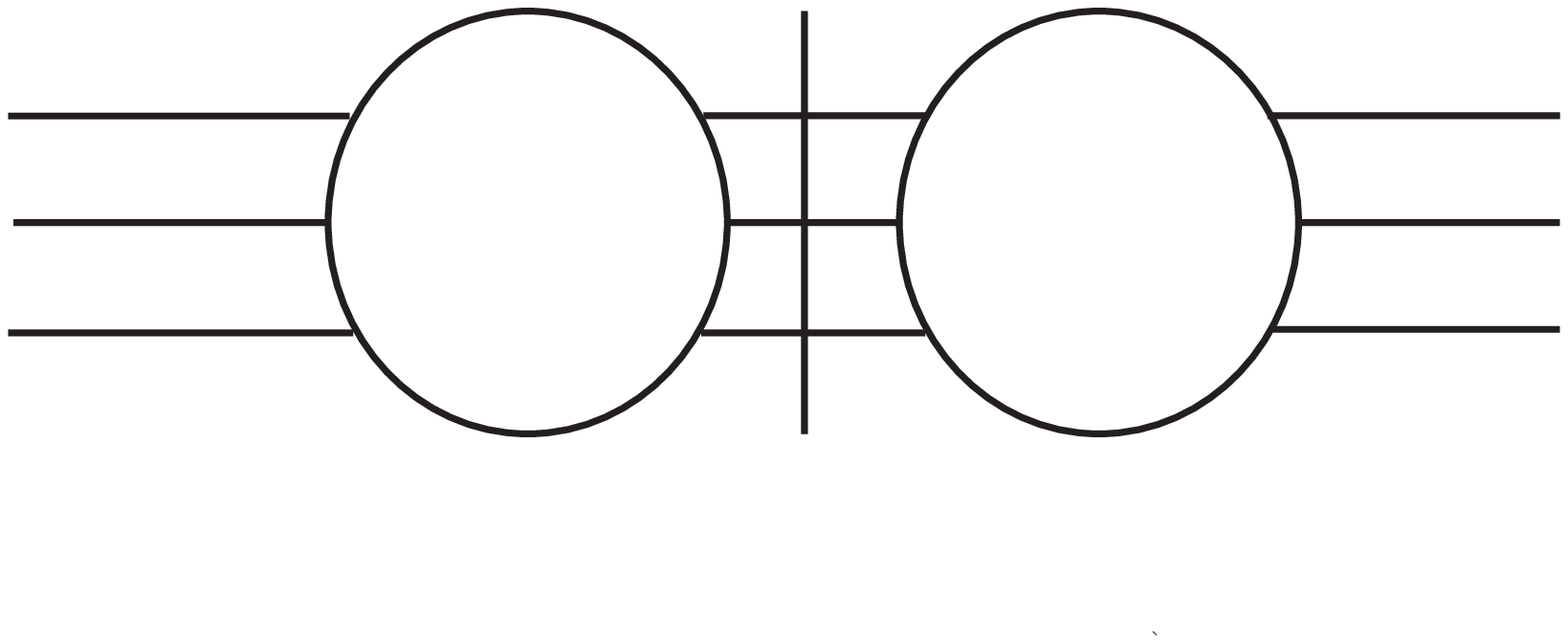}\end{center}
\vskip 0.1 in
\begin{center}Fig. 1.\end{center}
\vskip 0.1 in
This gives
\begin{equation}
A=\int...\int dk_1...dk_rA_I(q_j,k_i)\prod^r_{n=1}\frac{W_n(k_n^2,q)}{k_n^2+m_n^2}
A_{II}(q'_j,k_i)\delta^4(q-k_1-...-k_r),
\end{equation}
where $q_j$ and $q'_j$ denote the external momenta corresponding to the parts I and
II, respectively. The amplitude considered as a function of the variable $z=-q^2$ has
a branch cut starting at the point $z=(m_1+...+m_r)^2$ and the discontinuity of $A$
is given by \begin{equation} \label{Cutkosky} \Delta A(z)
=i(2\pi)^r\prod^r_{n=1}W_n(-m_r^2,q)\int d^4{\tilde k}_1...\int d^4{\tilde k}_r
\prod_{n=1}^r\theta({\tilde k}_{n0})\delta({\tilde k}_n^2+m_n^2) $$ $$
\times\delta^4({\tilde q}-{\tilde k}_1-... -{\tilde k}_r)A_I(q_j,{\tilde
k}_i)A_{II}(q'_j,{\tilde k}_i). \end{equation} Here, ${\tilde k}_i$ denotes 4-vectors
with the components $({\bf k},ik_{i0})$, with ${\tilde k}_i^2={\bf k}_i^2-k^2_{i0}$,
$({\tilde k}_iq_j)={\bf k}_i{\bf q}_j+i{\bf k}_{i0}q_{j4}$. The vector $q$ with
components $({\bf q},iq_0)$ satisfies $q^2= {\bf q}^2-q_0^2=-z$.

The result (\ref{Cutkosky}) is known as the Cutkosky cutting rule for normal
thresholds. It is satisfied by standard, local field theory perturbative amplitudes,
and can be extended to anomalous thresholds as well. The proof that it is satisfied
for amplitudes containing entire functions $W$ was given by Efimov~\cite{Efimov},
in the context of a nonlocal field theory,
using the methods of Landau~\cite{Landau} and Rudik and Simonov~\cite{Rudik}. The
proof consists in showing that it is possible to analytically continue the amplitude
$A(z)$ from the region $z<0$ to the physical region $z>0$ and to reveal the
singularities of the amplitude, which depend only on the intermediate momentum lines
of the diagram. The derivation of the proof for two internal lines leads to the
result
\begin{equation}
\Delta
A(z)=-(2\pi)^2\theta[z-(m_1+m_2)^2]W_1(-m_1^2,z)W_2(-m_2^2,z)C(-m_1^2,-m_2^2,z),
\end{equation}
where
\begin{equation}
\label{Cfunction}
C(-m^2_1,-m^2_2,z)
=\frac{-i[-u(m^2_1,m^2_2,z)]^{1/2}}{8z}
$$ $$
\times\int d\Omega_n
\Phi\biggl\{{\bf
n}\biggl[\frac{-u(m^2_1,m^2_2,z)}{4z}\biggr]^{1/2}\frac{i(m^2_1-m^2_2+z)}
{2\sqrt{z}}\biggr\}.
\end{equation}
Moreover, $z=-q^2$ and
\begin{equation}
u(a,b,c)=2ab+2bc+2ca-a^2-b^2-c^2 =[(\sqrt{a}+\sqrt{b})^2-c][c-(\sqrt{a}-\sqrt{b})^2].
\end{equation}
The integration in (\ref{Cfunction}) is performed with respect to all
the directions of the three-dimensional unit vector ${\bf n}$, and the ${\tilde q}$
vector has only a fourth component $({\bf q}=0,q_4=\sqrt{z})$. The proof of
(\ref{Cutkosky}) can be extended to diagrams with an arbitrary number of internal
lines.

It follows from these results that amplitudes calculated in the non-anticommutative
scalar field theory will not change the analyticity properties of the theory in any
finite region of the momentum variables, and the validity of the Cutkosky cutting
rules leads to a proof of the unitarity of this field theory. The singularities of
the amplitude in the momentum variables come only from the zeros of the denominator.
From this, it follows that the Landau and Cutkosky analyses can be applied and all
necessary conditions for the fulfillment of unitarity are satisfied. Landau required
that the amplitudes were regular at infinity in the momentum plane. He was then able
to perform a Wick contour rotation ($k_0\rightarrow ik_4$) and relate the zeros of a
diagram to the zeros of a known quadratic form in the external momenta subsequent to
an application of the Feynman $\alpha$ parameterization.

In Euclidean momentum space,
our amplitudes are not affected by possible essential singularities at infinity and
the positions of singularities of diagrams are likewise not affected. When we
analytically continue to Minkowski spacetime, all integrals converge and the
amplitudes are regular at infinity for $\tau^{00}=\tau^{0n}=0$ and $\tau^{mn}\not=
0\quad (m,n=1,2,3)$. This property of the amplitudes will be proved in the next
Section.

\section{Asymptotic Behavior of Amplitudes}

We shall now investigate the high energy behavior of amplitudes. Let us consider
the second order amplitudes described by diagrams of the type Fig. 2.
\vskip 0.3 true in
\begin{center}\includegraphics[width=3in,height=1in]{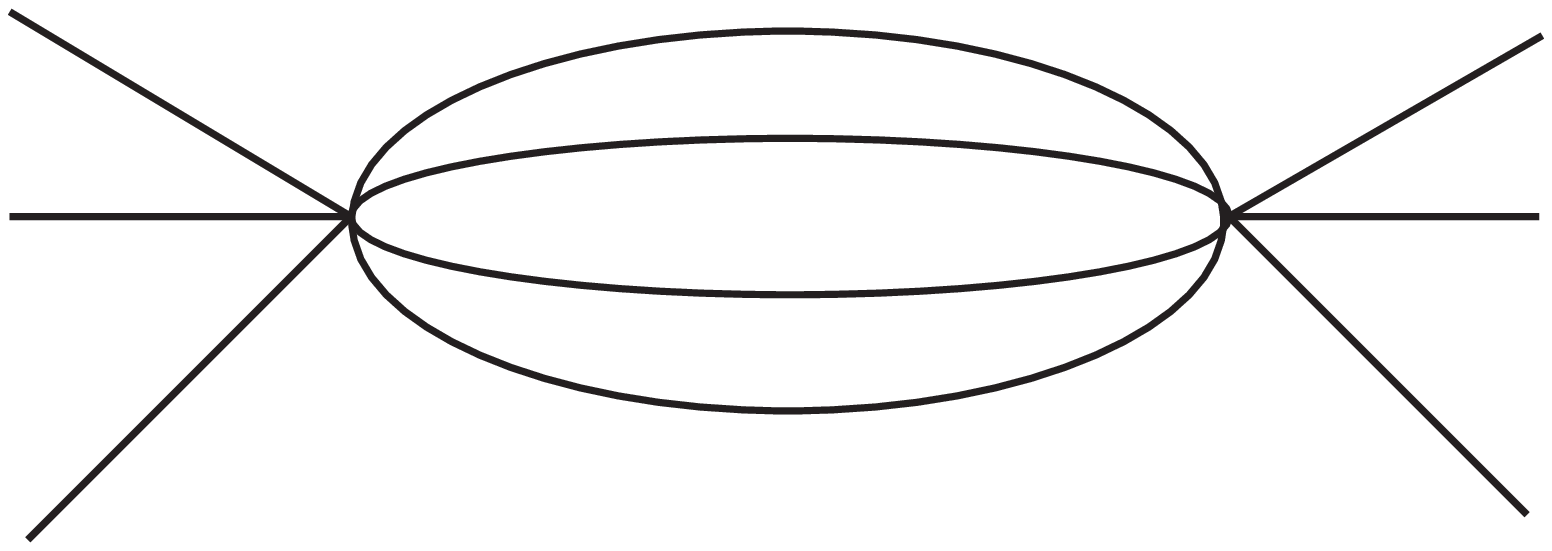}\end{center}
\vskip 0.1 in
\begin{center}Fig. 2\end{center}
\vskip 0.1 in
There are $r$ internal lines
and $n$ external lines meet at each vertex. We denote by $p$ the sum of the $n$
external momenta. The amplitude corresponding to this graph has the
form
\begin{equation}
A_2(p^2)=\lambda^2\int
d^4\rho\exp(ip\rho){\bar\Delta}^r_{mF}(\rho)V^r(\rho)
=\lambda^2\int d^4\rho\exp(ip\rho)W^r(\rho),
\end{equation}
where
${\bar\Delta}^r_{mF}$ is the modified causal propagator for a
mass $m$ and $r$ internal lines, $V^r(\rho)$ is the
non-anticommutative vertex factor and the integral is taken over
the four-dimensional superspace coordinates.

We use the relation
\begin{equation}
{\bar\Delta}^r_{mF}(\rho)=\int^\infty_{(mr)^2}
d\kappa^2\Omega^r(\kappa^2) {\bar\Delta}_{\kappa F}(\rho),
\end{equation} where $\Omega^r(\kappa^2)$ is the phase volume of
$r$ scalar particles with mass $m$:
\begin{equation}
\Omega^r(\kappa^2)=\frac{1}{(2\pi)^{3(r-1)}}\int\frac{d{\bf
k}_1}{2\omega_1}... \int\frac{d{\bf
k}_r}{2\omega_r}\delta^4(k-k_1-...-k_r), \end{equation} where
$\omega_i=\sqrt{{\bf k}_i^2+m^2}$ and on shell $k^2=k_0^2-{\bf
k}^2=\kappa^2$.

We now have
\begin{equation}
A_2(p^2)
=\lambda^2\int^\infty_{(mr)^2}d\kappa^2\Omega^r(\kappa^2){\bar\Delta}_\kappa(p^2)
V_\kappa(p)
$$ $$
=\lambda^2\int^\infty_{(mr)^2}d\kappa^2\Omega^r(\kappa^2)\biggl[\frac{B_\kappa(p)}{-p^2
+\kappa^2-i\epsilon}\biggr],
\end{equation}
where
\begin{equation}
\label{B(p)}
B_\kappa(p)=\exp\biggl[\frac{1}{2}(p\tau p)\biggr]V_\kappa(p).
\end{equation}

Writing $s=p^2$, the amplitude $A_2(s)$ is real in the region $s<(rm)^2$ and vanishes
for $s\rightarrow-\infty$. In the complex s-plane, it has a branch cut beginning at
$s=(rm)^2$, and the imaginary part is
\begin{equation}
{\rm Im}A_2(s)=\pi\lambda^2\Omega^r(s)B(\kappa^2)
\end{equation}
which is required by unitarity. We also have
\begin{equation}
\label{ReAmplitude}
{\rm Re}A_2(s)=\lambda^2\int^\infty_{(mr)^2}d\kappa^2\Omega^r(\kappa^2)
\biggl[\frac{B_\kappa(p)}{-s+\kappa^2}\biggr].
\end{equation}

The dominant behavior of $B_\kappa(p)$ as $p^2\rightarrow+\infty$ is
\begin{equation}
B_\kappa(p)\sim \exp[a(p\tau p)],
\end{equation}
where $a$ is a positive constant and $(p\tau p)=p_\mu\tau^{\mu\nu}p_\nu$. Let us
choose an orthonormal frame such that $\tau^{\mu\nu}=\eta^{\mu\nu}/\Lambda^2$ where
$\eta^{\mu\nu}={\rm diag}(+1,-1,-1,-1)$.
Then, for $p^2\rightarrow+\infty$, we have \begin{equation} B_\kappa(p)\sim
\exp\biggl(ap^2/\Lambda^2\biggr),
\end{equation}
and from (\ref{ReAmplitude}), it
follows that ${\rm Re}A_2(s)$ increases more rapidly than any finite power of $s$ as
$s\rightarrow+\infty$.

Let us instead choose an orthonormal frame with $\tau^{00}=\tau^{0n}=0$ and
$\tau^{mn}=-\delta_{mn}/\Lambda^2\quad (m,n=1,2,3)$. Then, we have
\begin{equation}
B_\kappa(p)\sim \exp\biggl(-a{\bf p}^2/\Lambda^2\biggr).
\end{equation}
Since ${\bf p}^2 >0$, it follows that for $s\rightarrow\pm\infty$ (and ${\bf
p}^2\rightarrow +\infty$), ${\rm Re}A_2(s)$ vanishes rapidly for $s\gg\Lambda^2$.

If we allow $\tau^{00}\not= 0$ and $\tau^{0i}\not= 0$, then ${\rm Re}A_2(s)$ will
have an essential singularity at $s\rightarrow+\infty$. While this will not violate
unitarity, because there are no additional singularities in the finite complex
$s$-plane, it will lead to unphysical behavior of the scattering amplitudes at high
energies and unphysical crossing symmetry relations. However, our choice of
$\tau^{\mu\nu}$ {\it breaks Lorentz invariance}. It was argued by
Efimov~\cite{Efimov} that a summation of a certain class of graphs can correct the
high energy behavior problem of the amplitudes. But there is no convincing proof that
the unphysical increase of amplitudes can be compensated for by the inclusion of
higher order graphs.

\section{\bf Conclusions}

We have investigated the Cutkosky rules and unitarity in a
non-anticommutative scalar field theory. For arbitrary Feynman diagrams in Euclidean
space, the Cutkosky rules are valid for normal thresholds. There are no additional
unphysical singularities present in the complex momentum plane and the transition to
the physical region of external momenta can be achieved by analytic continuation of
the amplitudes with respect to the invariant momenta. The modified Feynman
propagator ${\bar\Delta}_F(p^2)$ and the vertex factor $V(k,p)$ are entire functions
of the momenta, which allows an analysis of the singularity structure of the
amplitudes in perturbation theory in the manner of Landau and Efimov. The results
lead to a proof of unitarity.

The high energy behavior of scattering amplitudes was studied and it was found that
crossing symmetry relations and high energy behavior can be physical, provided that
we restrict ourselves to the case $\tau^{mn}\not= 0$ and
$\tau^{00}=\tau^{0n}=0$. The same circumstances exist for noncommutative field
theories in which $[{\hat x}^\mu,{\hat x^\nu}]=i\theta^{\mu\nu}$, because the
restriction that $\theta^{mn}\not= 0$ and $\theta^{0n}=0$ avoids problems with
unitarity and causality. We can avoid a `hard' breaking of Lorentz invariance by
invoking a spontaneous breaking of Lorentz invariance~\cite{Moffat2,Moffat3}. In
this scenario, a Higgs vector symmetry breaking mechanism is introduced which allows
a soft breaking of $SO(3,1)$ to $O(3)$ at short distance scales when the
non-anticommutative natural unit of length $\ell\not= 0$ and
$\vert\tau^{\mu\nu}\vert\not=0$.

\vskip 1 true in {\bf
Acknowledgments}
\vskip 0.2 true in
I thank Michael Clayton for helpful discussions. This work was supported by the
Natural Sciences and Engineering Research Council of Canada.
\vskip 0.5 true in

\end{document}